\documentclass[12pt]{article}
\usepackage{amsmath,amssymb,epsfig}
\usepackage{overpic}
\usepackage{graphicx,floatflt,subfigure}

\usepackage{color}
\input{colordvi.tex}

\renewcommand{\thefootnote}{\fnsymbol{footnote}}

\makeatletter

\renewcommand\section{\@startsection {section}{1}{\z@}%
                                   {-3.5ex \@plus -1ex \@minus -.2ex}%
                                   {2.3ex \@plus.2ex}%
                                   {\normalfont\large\bfseries}}

\renewcommand\subsection{\@startsection{subsection}{2}{\z@}%
                                     {-3.25ex\@plus -1ex \@minus -.2ex}%
                                     {1.5ex \@plus .2ex}%
                                     {\normalfont\normalsize\bfseries}}

\newcount\hour \newcount\minute
\hour=\time \divide \hour by 60
\minute=\time
\def\now{%
\ifnum \hour<13
  \ifnum \hour=0 \advance \hour by 12 \number\hour:\else \number\hour:\fi%
     \ifnum \minute<10 0\fi%
     \number\minute%
\ A.M.%
\else \advance \hour by -12 \number\hour:%
  \ifnum \minute<10 0\fi%
  \number\minute%
  \ P.M.%
\fi%
}

\makeatother

\begin{document}

\baselineskip=18pt  
\numberwithin{equation}{section}  
\allowdisplaybreaks  



%
%


\thispagestyle{empty}

\vspace*{-2cm}
\begin{flushright}
\end{flushright}

\begin{flushright}
RESCEU-5/11 \\
YITP-11-36 \\
\end{flushright}

\begin{center}

\vspace{2.0cm}

{\bf\large Inflation in Gauge Mediation and Gravitino Dark Matter}
\\

\vspace*{1.5cm}
{\bf
Kohei Kamada$^{1,2}$\footnote{e-mail: {\tt kamada\_at\_resceu.s.u-tokyo.ac.jp}},
Yuichiro Nakai$^{\,3}$\footnote{e-mail:
{\tt ynakai\_at\_yukawa.kyoto-u.ac.jp}} and Manabu Sakai$^{\,3}$}\footnote{e-mail: {\tt msakai\_at\_yukawa.kyoto-u.ac.jp}} \\
\vspace*{0.8cm}

$^{1}${\it Department of Physics, Graduate School of Science, \\
The University of Tokyo, Tokyo 113-0033, Japan} \\
\vspace{0.1cm}

$^{2}${\it Research Center for the Early Universe (RESCEU), \\
Graduate School of Science, The University of Tokyo, Tokyo 113-0033, Japan} \\
\vspace{0.1cm}

$^{3}${\it {Yukawa Institute for Theoretical Physics, Kyoto University,
   Kyoto 606-8502, Japan}}
\vspace{0.1cm}

\end{center}

\vspace{1.3cm} \centerline{\bf Abstract} \vspace*{0.5cm}

We present an inflationary scenario based on a phenomenologically viable model with direct gauge mediation of low-scale supersymmetry breaking. 
Inflation can occur in the supersymmetry-breaking {hidden} sector. 
Although the reheating temperature from the inflaton decay is so high that the gravitino problem seems to be severe, 
late time entropy production from the decay of the pseudomoduli field associated with the supersymmetry breaking 
can dilute gravitinos {sufficiently}. 
We show that gravitinos are also produced from the pseudomoduli decay and 
there is a model parameter space where gravitinos can be the dark matter in the present universe.


\newpage
\setcounter{footnote}{0}
\renewcommand{\thefootnote}{\arabic{footnote}}





\section{Introduction}

Cosmological inflation \cite{Sato:1980yn,infrev} in the early universe is now considered 
as a part of the ``standard" cosmology. It can solve many cosmological problems
such as the horizon problem and the flatness problem. It also accounts 
for the origin of the primordial fluctuations \cite{fluctuation}. 
In most models of inflation, a scalar field with the flat potential (called the inflaton) which drives inflation is assumed 
to be outside the standard model of particle physics since there is almost no inflaton candidate 
in this framework.\footnote{It is usually considered that Higgs field cannot be responsible for inflation since
its potential is too steep to account for the primordial fluctuations \cite{Linde:1983gd}.
Recently, however, several inflationary models that regard the standard model Higgs boson as the inflaton 
were proposed \cite{higgsinf}. In these models, the Higgs self-coupling is effectively suppressed by increasing 
the effective Planck mass or the Higgs kinetic term and hence primordial fluctuations consistent with the 
observational data \cite{Komatsu:2010fb} can be generated.} Then, a natural question is how inflation dynamics 
is embedded in a particle physics model beyond the standard model. 
To construct an inflation model, it {is} reasonable to respect supersymmetry (SUSY), 
which is one of the most promising candidates for physics beyond the standard model, 
because the radiative corrections for scalar fields are suppressed in supersymmetric theories 
and an inflaton potential can be naturally flattened.

If SUSY is realized in nature, it must be broken at some energy scale. 
We usually leave its dynamics to a SUSY breaking {hidden} sector and the SUSY breaking effect is 
transmitted to the visible sector by some interactions. Gauge mediation is an attractive mechanism,
which uses the standard model gauge interactions \cite{GaugeMediation,Giudice:1998bp} 
(For recent general arguments, see Ref.~\cite{Meade:2008wd}). 
It naturally suppresses the unwanted flavor-changing processes due to flavor blindness of the gauge interactions. 
While many gauge mediation models are already known, the model proposed by Kitano, Ooguri and Ookouchi (KOO) 
\cite{Kitano:2006xg} is a distinguished one (see also Ref.~\cite{Giveon:2009yu}). 
This model is based on the recent development of metastable SUSY breaking in supersymmetric QCD (SQCD) started 
from the work of Intriligator, Seiberg and Shih (ISS) \cite{Intriligator:2006dd} 
and is classified as so-called direct gauge mediation models \cite{Izawa:1997gs} where the flavor symmetries of 
a hidden SUSY breaking sector are weakly gauged and identified with the standard model gauge symmetries. 
The KOO model can generate sizable gaugino masses while many direct gauge mediation models suffer from 
anomalously small gaugino masses \cite{Komargodski:2009jf} (see also Ref.~\cite{Nakai:2010th}). 
However, since the vacuum considered in the KOO model (the ISS vacuum) is not the global minimum of the potential, 
it is necessary to impose several conditions on the model parameters in order to guarantee the vacuum stability. 
In particular, two hierarchical mass scales are required in the flavors of this massive SQCD model. 
We must also consider the cosmic history that selects the ISS vacuum in this model.\footnote{One approach is to consider 
the finite temperature effect on the SUSY breaking sector \cite{finite}. 
For instance, it is known that the metastable vacuum is preferred than the SUSY preserving vacuum 
at the high temperature in the ISS model. However, this approach has a difficulty that gravitinos are 
overproduced when the SUSY breaking sector is thermalized.}  

Recently, two of the present authors  realized hybrid inflation \cite{hybrid} embedded in a gauge-mediated 
SUSY-breaking model motivated by the KOO model \cite{Nakai:2010km}.
This scenario uses two hierarchical mass scales in the magnetic description of the KOO model 
such that the higher scale corresponds to the inflationary scale and the lower scale corresponds 
to the SUSY breaking scale. Since the inflaton rolls down to the ISS vacuum after inflation, 
it naturally explains why the SUSY breaking vacuum is selected in the cosmic history. 
Another attractive feature of this model is that if we can find the model parameter space that accommodates
the overall cosmic history (including inflation and the present dark matter abundance) and gauge mediation, 
we might be able to prove (or disprove) the scenario by the near future collider experiments and 
the cosmological observations such as Large Hadron Collider (LHC) and PLANCK \cite{Planck}.

In this paper, we present an inflationary scenario following the line of Ref.~\cite{Nakai:2010km} and analyze the reheating stage of an inflation model 
with gauge mediation. Since the couplings of the inflaton in the SUSY breaking hidden sector with the visible sector fields are given by the standard model gauge interactions, the reheating process after inflation is predictable in this framework. We here take a simpler Wess-Zumino model than that in Ref.~\cite{Nakai:2010km} as the SUSY breaking hidden sector, but 
the essential feature is the same.
As is the case in Ref.~\cite{Ibe:2006rc} (see also Ref.~\cite{Endo:2007sw}), the evolution of the universe after 
inflation proceeds as follows. 
{After the end of (hybrid) inflation, the damped oscillation of the inflaton and waterfall fields dominate 
the energy density of the universe and the Hubble parameter of the universe decreases.}
At the stage of reheating from the inflaton decay through the messenger loops, only the fields in the visible sector are thermalized, 
since {those} in the SUSY breaking sector are heavy.
As the {Hubble parameter} decreases, the pseudomoduli associated with the SUSY breaking 
that was stabilized at the origin of the field space during inflation starts to oscillate around the metastable 
vacuum.\footnote{It depends on the model parameters that the moduli oscillation takes place earlier or later than reheating.} 
Because the lifetime of the pseudomoduli field is rather long, the pseudomoduli oscillation dominates the universe.  
Then, the pseudomoduli decays before the Big Bang Nucleosynthesis (BBN). 
Gravitinos produced in the thermal bath at the reheating stage are diluted enough by the entropy production 
from its decay, whose present abundance otherwise tends to be so large that it overcloses the universe.
Although gravitinos are also produced in this decay process, the branching fraction is suppressed 
since the pseudomoduli can decay into the visible sector particles through the messenger loops. 
Thermally and non-thermally produced gravitinos can both contribute the dark matter abundance 
in the present universe.


The rest of the paper is organized as follows. In section 2, we present our inflation model and analyze the vacuum 
structure and the mass spectrum around the metastable vacuum. 
We also evaluate the soft mass spectrum of the visible sector fields from gauge mediation. 
In section 3, we present the inflationary scenario based on Ref.~\cite{Nakai:2010km} 
and estimate the reheating temperature after inflation. In section 4, the oscillation of the pseudomoduli field is analyzed. 
Next, we investigate the decay temperature of the pseudomoduli field. 
We will also show the gravitino abundance due to the pseudomoduli decay. 
Then, we find the model parameter space consistent with the BBN constraints and the present dark matter abundance. 
In section 6, we conclude our discussions.

\section{SUSY breaking}

In this section, we present a metastable SUSY breaking model, which drives inflation and gauge mediation, 
and analyze its vacuum structure. Then, we analyze the mass spectrum around the SUSY breaking vacuum. 
We also show the soft mass spectrum of the visible sector fields obtained by gauge mediation.

\subsection{The model}

\begin{table}[tdp]
\begin{center}
\begin{tabular}{c|cccccc}
 & $SU(N)$ & $U(1)_1$ & $U(1)_2$ & $U(1)_R$
 \\
 \hline
$\chi$ & $\mathbf{1}$ 
& $1$ & $0$ & $0$
\\
$\bar{\chi}$ & $\mathbf{1}$
& $-1$ & $0$ & $0$
\\
$\rho$  & $\square$ 
& $0$ & $1$ & $0$
  \\
$\bar{\rho}$  & $\bar{\square}$ 
& $0$ & $-1$ & $0$
  \\
$Z$ & $\square$ & $-1$ & $1$ & $2$ \\
$\bar{Z}$ & $\bar{\square}$ & $1$ & $-1$ & $2$ \\
  $Y$ & $\mathbf{1}$ & $0$ & $0$ & $2$ \\
  $\Phi$ & $\mathbf 1$  & $0$ & $0$ & $2$
\end{tabular}
\end{center}
\caption{The charge assignments of the fields in our model.}
\label{tab:charge}
\end{table}

The model is a Wess-Zumino model with an $SU(N)$ global symmetry.\footnote{Later, we gauge this global symmetry 
to be the standard model gauge symmetry.} 
The matter content is summarized in Table~\ref{tab:charge}. 
$\chi$ and  $\bar{\chi}$ are singlets under the $SU(N)$ group and the fields $\rho$, $\bar{\rho}$ 
belong to the (anti-)fundamental representation under the $SU(N)$ group. 
$Z$ and $\bar{Z}$ are also a vector-like pair of the (anti-)fundamental representation 
under the $SU(N)$ group. 
$Y$ and $\Phi$ are singlets under all the global symmetries. 
The tree-level K\"ahler potential of all fields is assumed to be canonical. 
The superpotential is given by
\begin{equation}\label{dualsuperpotential}
W = m^2 Y + \mu^2 \Phi - h_{\rm Y} \chi Y \bar{\chi} - h_{\rm \Phi} \rho \Phi \bar{\rho} - h_{\rm Z} ( \chi Z \bar{\rho} + \rho \bar{Z} \bar{\chi} ) - m_{\rm Z} Z \bar{Z},
\end{equation}
where the mass scale $m$ is assumed to be larger than the scale $\mu$, and $h_{\rm Y}$, $h_{\rm \Phi}$, 
$h_{\rm Z}$ are coupling constants. 
The model has two $U(1)$ global symmetries. $U(1)_{R}$ symmetry is explicitly broken by the last term in the 
superpotential, whose breaking size is determined by the mass scale $m_{\rm Z}$. 
First, we treat them as free parameters and will determine their sizes later as they pass all the conditions.
For simplicity, we assume that all the couplings and the mass parameters are real.\footnote{The sizes of the mass 
parameters can be explained by introducing additional gauge dynamics \cite{Dine:2006gm}.}

Let us analyze the vacuum structure of our model. 
There exists a metastable SUSY breaking vacuum\footnote{In the absence of the last term in 
\eqref{dualsuperpotential}, SUSY is always broken in the model.} 
where the expectation values of the fields are denoted as\footnote{In order to obtain 
the equal expectation values of the $\chi$, $\bar{\chi}$ fields, we have used the additional condition from 
the D-term potential when we gauge the $U(1)_1$ symmetry.}
\begin{equation}
Y = \rho = \bar{\rho} = Z = \bar{Z} = 0
, \quad
\chi = \bar{\chi} =
\frac{m}{\sqrt{h_{\rm Y}}},
\end{equation}
and the singlet $\Phi$ is the pseudomoduli of the SUSY breaking vacuum that is massless at the tree-level, 
but obtains a nonzero mass from radiative corrections. 
Since the vacuum energy is given by $V_{0} = \mu^4$ at the tree-level, the gravitino mass is estimated as
\begin{equation}
m_{3/2} = \frac{\mu^2}{\sqrt{3}M_{\rm Pl}},
\end{equation}
where $M_{\rm Pl} \simeq 2.43 \times 10^{18} $ GeV is the reduced Planck mass. Conversely, when we express the SUSY breaking order parameter $\mu$ in terms of the gravitino mass, it is given by
\begin{equation}
\mu \simeq 7.9 \times 10^9 \, \text{GeV} \times \left( \frac{m_{3/2}}{15 \, \text{GeV}} \right)^{1/2},
\label{mugrav}
\end{equation}
where we take $15 \, \text{GeV}$ for the gravitino mass as the reference value.

The pseudomoduli $\Phi$ is stabilized by the 1-loop effects of the massive modes $\rho$, $\bar{\rho}$, 
$Z$ and $\bar{Z}$. The supersymmetric mass terms for these massive modes are given by the following mass matrix:
\begin{equation}
( \rho , Z)
M
\begin{pmatrix}
\bar{\rho}  \\
\bar{Z}
\end{pmatrix} =
( \rho , Z)
\begin{pmatrix}
- h_{\rm \Phi} \Phi & -\frac{h_{\rm Z}}{\sqrt{h_{\rm Y}}}m \\
- \frac{h_{\rm Z}}{\sqrt{h_{\rm Y}}}m & -m_{\rm Z}
\end{pmatrix}
\begin{pmatrix}
\bar{\rho}  \\
\bar{Z}
\end{pmatrix}.
\label{massmatrix}
\end{equation}
Integrating out these modes, we can derive the 1-loop effective K\"ahler potential for the pseudomoduli field $\Phi$ 
as \cite{Grisaru:1996ve}
\begin{equation}
K_{\rm eff} \simeq |\Phi|^2 - \frac{N}{32\pi^2} \mathrm{Tr} \left[ M M^\dagger \left( \log{\frac{M M^\dagger}{\Lambda^2}} - 1 \right) \right],
\end{equation}
where $\Lambda$ is a cut-off scale. Substituting the expression of the mass matrix \eqref{massmatrix} into this formula, 
we can obtain the following effective K\"ahler potential at the leading order of the scale $m_{\rm Z}$:
\begin{equation}
\begin{split}
K_{\rm eff}
&\simeq |\Phi|^2 -\frac{N}{32\pi^2} \biggl[ h_{\rm \Phi} m_{\rm Z} \left(\Phi + \Phi^\dagger \right) + h_{\rm \Phi}^2 |\Phi|^2 \\
&\quad - \frac{1}{8} \frac{h_{\rm Y} h_{\rm \Phi}^3}{h_{\rm Z}^2} \frac{m_{\rm Z}}{m^2} |\Phi|^2 \left(\Phi+\Phi^\dagger \right) + \frac{1}{8}\frac{h_{\rm Y}h_{\rm \Phi}^4}{h_{\rm Z}^2} \frac{1}{m^2} |\Phi|^4 + \mathcal{O}(m_{\rm Z}^2)\biggr].
\end{split} \label{keffmod}
\end{equation}
Note that the $U(1)_R$ breaking terms in the above expression are proportional to the mass parameter $m_{\rm Z}$. 
The 1-loop effective scalar potential for the pseudomoduli field $\Phi$ can be derived from the effective 
K\"ahler potential as
\begin{equation}
V_{\rm eff} \simeq \left( \frac{\partial^2 K_{\rm eff}}{\partial\Phi \partial \Phi^\dagger} \right)^{-1} \left| \frac{\partial W}{\partial \Phi} \right|^2.
\end{equation}
Then, we can extract the vacuum expectation value and the mass squared of the pseudomoduli field $\Phi$ from 
the above effective scalar potential, which are given by
\begin{equation}
\begin{split}
|\Phi_0| &\simeq \frac{1}{2} \frac{m_{\rm Z}}{h_{\rm \Phi}}, \quad \arg \Phi_0=0, \\
\quad \ m^2_\Phi &\simeq \frac{N}{64\pi^2} \frac{h_{\rm Y}h_{\rm \Phi}^4}{h_{\rm Z}^2 } \frac{\mu^4}{m^2} \equiv m_{\rm CW}^2.
\end{split}
\label{pseudomass}
\end{equation}
As we will see, the field configurations finally settle down to this vacuum after inflation. 
Later, we consider the gauge mediation effects on the visible sector fields at this vacuum.

There is a SUSY preserving vacuum far away from the origin of the pseudomoduli field space, 
in addition to the metastable vacuum. 
The vacuum expectation values of the fields are given by
\begin{equation}
\begin{split}
&\chi \bar{\chi} = \frac{m^2}{h_{\rm Y}}, \quad \rho \bar{\rho} = \frac{\mu^2}{h_{\rm \Phi}}, \quad Z \bar{Z} = \frac{h_{\rm Z}^2}{h_{\rm Y}h_{\rm \Phi}} \frac{m^2\mu^2}{m_{\rm Z}^2},\\
&Y = \frac{h_{\rm Z}^2}{h_{\rm Y}h_{\rm \Phi}} \frac{\mu^2}{m_{\rm Z}}, \quad \Phi = \frac{h_{\rm Z}^2}{h_{\rm Y}h_{\rm \Phi}} \frac{m^2}{m_{\rm Z}}.
\end{split}
\end{equation}
The SUSY breaking vacuum has a nonzero transition probability to the SUSY preserving vacuum, which can destroy 
the stability of the SUSY breaking vacuum and hence SUSY breaking mechanism may not work. 
However, its decay rate $\Gamma_{\rm vac}$ is evaluated by using the triangle approximation as \cite{Duncan:1992ai}
\begin{align}
\Gamma_{\rm vac}&\propto e^{-S}, \\
S &\sim \frac{h_Z^8}{h_Y^4 h_\Phi^4}\left(\frac{m}{m_Z}\right)^4\left( \frac{m}{\mu} \right)^4.
\end{align}
Then, if the mass hierarchies $m \gg \mu$ and $ m \gg m_Z$ are realized, we have $S\gg 1$.  
Thus, the decay rate is sufficiently suppressed and hence the stability of the metastable vacuum is guaranteed.


\subsection{Mass spectrum}

\begin{table}[t]
\begin{center}
{\scalebox{0.8}
{\renewcommand\arraystretch{2.5}
  \begin{tabular}{|c||ccc|ccc|} 
    \hline
      & \multicolumn{3}{|c|}{Fermions} & \multicolumn{3}{|c|}{Bosons} \\
    \hline
      & Weyl mult. & mass  & $SU(N)$ & Real mult. & mass  & $SU(N)$ \\
    \hline
    $\Phi$ & $1$ & 0 & $\mathbf{1}$ & $2$ & $m_{\rm CW}$ & $\mathbf{1}$ \\
    \hline
    $Y$, $\chi$, $\bar{\chi}$ & $1$ & $\mathcal{O}( \sqrt{h_{\rm Y}}m)$ & $\mathbf{1}$ & $2$ & $\mathcal{O}( \sqrt{h_{\rm Y}}m)$ & $\mathbf{1}$ \\
                              & $1$ & $\mathcal{O}( \sqrt{h_{\rm Y}}m)$ & $\mathbf{1}$ & $2$ & $\mathcal{O}( \sqrt{h_{\rm Y}}m)$ & $\mathbf{1}$ \\
                            
                              & $1$ & $g_{\rm V}\frac{m}{\sqrt{h_{\rm Y}}}$ & $\mathbf{1}$ & $2$ & $g_{\rm V}\frac{m}{\sqrt{h_{\rm Y}}}$ & $\mathbf{1}$ \\
    \hline
    $Z$, $\bar{Z}$, $\rho$, $\bar{\rho}$ & $2N$ & $\mathcal{O}( \frac{h_{\rm Z}}{\sqrt{h_{\rm Y}}}m)$ & $\square + \bar{\square}$ & $4N$ & $\mathcal{O}( \frac{h_{\rm Z}}{\sqrt{h_{\rm Y}}}m)$ & $\square + \bar{\square}$ \\
                                         & $2N$ & $\mathcal{O}( \frac{h_{\rm Z}}{\sqrt{h_{\rm Y}}}m)$ & $\square + \bar{\square}$ & $4N$ & $\mathcal{O}( \frac{h_{\rm Z}}{\sqrt{h_{\rm Y}}}m)$ & $\square + \bar{\square}$ \\
    \hline
  \end{tabular}
}
}
\end{center}
\caption{The mass spectrum and the representations under the $SU(N)$ symmetry. In the table, $g_{\rm V}$ denotes the $U(1)_1$ gauge coupling. The scalar component of the pseudomoduli field $\Phi$ has a mass \eqref{pseudomass} by the 1-loop effects.}
\label{tab:mass}
\end{table}

We next analyze the mass spectrum around the SUSY breaking vacuum. 
The masses of the fields in the SUSY breaking sector except for the pseudomoduli are generated at the tree-level.
The results of the masses as well as the representations of the fields under the symmetries are summarized in 
Table~\ref{tab:mass}. 
The mass spectrum of the $Y$, $\chi$, $\bar{\chi}$ sector fields are supersymmetric at the tree-level. 
All of the scalar fields in this sector except the Nambu-Goldstone mode have $\mathcal{O}( \sqrt{h_{\rm Y}}m)$ masses. 
The corresponding fermionic modes have the same masses at the tree-level. 
The Nambu-Goldstone mode is associated with the spontaneous breaking of the $U(1)_1$ symmetry 
due to the nonzero expectation values of the $\chi$, $\bar{\chi}$ fields. 
In order to give this mode a nonzero mass, we can weakly gauge the $U(1)_1$ symmetry. 
Then, the Nambu-Goldstone mode is absorbed into the massive gauge boson of the symmetry group. 
When we gauge the spontaneous breaking global symmetry, the fermionic partner of the Nambu-Goldstone boson 
becomes a part of the massive vector multiplet.

The $Z$, $\bar{Z}$, $\rho$, $\bar{\rho}$ fields have the SUSY breaking mass spectrum 
and hence play the role of messengers in gauge mediation. 
The fermion mass matrix is given in \eqref{massmatrix} and they all have 
$\mathcal{O}( \frac{h_{\rm Z}}{\sqrt{h_{\rm Y}}}m)$ supersymmetric masses which correspond to the messenger scale
in our gauge mediation model. The SUSY breaking masses of the scalar components are small 
compared to the supersymmetric ones.

As shown in the previous subsection, the scalar component of the singlet field $\Phi$ obtains a nonzero mass 
due to the 1-loop effects. The fermion component corresponds to the would-be goldstino mode associated with the SUSY breaking, which is absorbed into the massive gravitino when the theory is promoted to supergravity.

\subsection{Gauge mediation}

We next consider the soft mass spectrum of the visible sector fields. 
In order to transmit the SUSY breaking of the metastable vacuum to the visible sector by direct gauge mediation, 
we embed the standard model gauge group into the $SU(N)$ symmetry of the model. 
The formulae of the leading order gaugino and scalar masses are given by \cite{Cheung:2007es}
\begin{equation}
\begin{split}
m_{\lambda_i} &\simeq \frac{g_i^2}{16\pi^2} \, F_\Phi \frac{\partial}{\partial \Phi} \log \det M, \\
m_{\tilde{f}}^2 &\simeq \sum_i C_2^i \left( \frac{g_i^2}{16\pi^2} \right)^2 |F_\Phi|^2 \frac{\partial^2}{\partial \Phi \partial \Phi^\dagger} \sum_s \left( \log \left| M_s \right|^2 \right)^2,
\end{split}\label{softformula}
\end{equation}
where $g_i \, (i=1,2,3)$ is the gauge coupling constant of the $U(1) \times SU(2) \times SU(3)$ group of the standard model 
gauge symmetry and the factor $C_2^i$ is a quadratic Casimir. 
$M$ is the mass matrix of the messenger fields $\rho$, $\bar{\rho}$, $Z$, and $\bar{Z}$ 
which is given in \eqref{massmatrix}. 
$M_s$ denotes the eigenvalue of this matrix. 
Then, at the lowest order of the scale $m_Z$, the gaugino and scalar masses in our model are explicitly given by
\begin{equation}
\begin{split}
m_{\lambda_i} &\simeq \frac{g_i^2}{16\pi^2} \, \frac{h_{\rm Y}h_{\rm \Phi}}{h_{\rm Z}^2} \, \frac{\mu^2}{m} \, \frac{m_{\rm Z}}{m}, \\
m_{\tilde{f}}^2 &\simeq \sum_i C_2^i \left( \frac{g_i^2}{16\pi^2} \right)^2 \frac{h_{\rm Y}h_{\rm \Phi}^2}{h_{\rm Z}^2} \, \frac{\mu^4}{m^2}.
\end{split}\label{soft}
\end{equation}
As we can see in the above expressions, the leading order gaugino masses do not vanish. 
Since the metastable vacuum is the higher-energy state in the potential, 
this result is consistent with the general theorem presented in \cite{Komargodski:2009jf}.
Now, we parametrize the ratio between the gluino mass $m_{\tilde{g}}$ and the right-handed slepton mass 
$m_{\tilde{e}}$ for later use as follows:
\begin{equation}
r_{\rm g} = m_{\tilde{g}} / m_{\tilde{e}}. 
\end{equation}
In our scenario, this ratio is constrained by the requirements to accommodate the proper inflation, BBN and dark matter abundance in the present universe.

We can rewrite the coupling $h_{\rm \Phi}$ in terms of the observable parameters by using the expressions of the moduli mass, the soft scalar mass, and the gaugino-to-scalar mass ratio $r_{\rm g}$ as follows:
\begin{equation}
h_{\rm \Phi} \simeq 0.036 \times \frac{1}{\sqrt{N}} \left( \frac{r_{\rm g}}{3.5} \right) \left( \frac{m_{\rm \Phi}}{300 \, \text{GeV}} \right) \left( \frac{m_{\tilde{g}}}{1.5 \, \text{TeV}} \right)^{-1}.
\label{h-phi}
\end{equation}
We can see from this expression that the coupling $h_{\rm \Phi}$ is an $\mathcal{O} (10^{-2})$ quantity when the observable mass parameters have the reference values.

\section{Inflationary scenario}

In this section, we study the hybrid inflation model embedded in the above theory. 
We see that the $Y$ scalar field acts as {the} inflaton and 
the $\chi$, ${\bar \chi}$ fields act as the waterfall fields. 
We {analyze} its dynamics and density fluctuation generated during inflation. 
We also study the decay of the inflationary fields after inflation. 

\subsection{Inflationary dynamics and cosmological perturbation}

First, we see the realization of hybrid inflation which can occur on the pseudo-flat direction (different from the pseudomoduli $\Phi$ of the metastable vacuum) in our model.
In order to do so, {we concentrate on the following field configuration of the system:} 
\begin{equation}
\rho = \bar{\rho} = Z = \bar{Z} = \Phi = 0, 
\label{trajectory}
\end{equation}
which respects the global symmetry of the model.
{The relevant part of the superpotential \eqref{dualsuperpotential} is, now,}
\begin{equation}
W \simeq m^2 Y -h_{\rm Y} \chi Y {\bar \chi}. 
\end{equation} 
This type of superpotential is exactly of the form for F-term hybrid inflation \cite{hybrid}.
That is, the field $Y$ is the inflaton and the fields $\chi$, ${\bar \chi}$ are the waterfall fields.
The tree-level scalar potential in the global SUSY limit is, then, given by
\begin{equation}
V_{\rm tree} \simeq \left| m^2 - h_{\rm Y} \chi \bar{\chi} \right|^2 + h_{\rm Y}^2 |Y|^2 ( |\chi|^2 + |\bar{\chi}|^2 ).
\end{equation}	
The masses of the scalar fields $\chi, {\bar \chi}$ are
$m_{s}^2= h_{\rm Y}^2|Y|^2\pm h_{\rm Y}m^2$.
For the large field value of $Y$, the fields $\chi$, ${\bar \chi}$ are stabilized at the origin of their field space and hybrid inflation with the Hubble parameter,
\begin{equation}
3H^2M_{\rm Pl}^2\simeq V_{\rm tree} \Leftrightarrow H\simeq  \sqrt{\frac{1}{3}}\frac{m^2}{M_{\rm Pl}}
\end{equation}
can occur.
On the other hand, for the small field value of $Y$, $|Y|<Y_c \equiv m/\sqrt{h_{\rm Y}}$, one of the scalar modes is
destabilized. At that time, $\chi$ and ${\bar \chi}$ act as the waterfall fields 
and roll off to the point near the metastable vacuum discussed in the previous section with nonzero 
vacuum expectation values. 

\begin{figure}[t]
 \begin{center}
  \includegraphics[width=100mm]{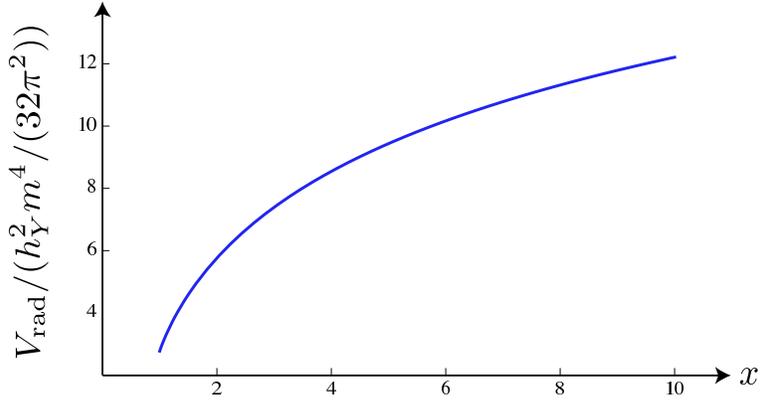}
 \end{center}
 \caption{The shape of the scalar potential $V_{\rm rad}$ is shown. 
It is convex upward and hence the slow-roll parameter $\eta$ will be negative. 
 }
 \label{fig:radpot} 
\end{figure}

Let us study the radiative correction, which drives the inflaton motion. 
Here, we redefine $Y\equiv \dfrac{y}{\sqrt{2}}e^{i\theta}$ and consider the motion of the inflaton $y$. 
The  potential for the phase $\theta$ remains flat and hence it does not have any influence on the inflaton dynamics.
The mass splittings between the bosonic and fermionic partners of $\chi$ and ${\bar \chi}$ due to the SUSY breaking 
result in the 1-loop effective potential for $y$ as
\begin{equation}
\begin{split}
V_{\rm rad}(y)
&=  \frac{1}{32\pi^2 } \Biggl[ 2 h_{\rm Y}^2 m^4 \log{\left( \frac{h_{\rm Y}^2 y^2}{2M_{\ast}^2} \right)} \\
&+ \left( \frac{h_{\rm Y}^2 y^2 }{2}+ h_{\rm Y}m^2 \right)^2 \log{\left( 1 + \frac{2m^2}{h_{\rm Y}y^2} \right)} + \left( \frac{h_{\rm Y}^2 y^2 }{2} - h_{\rm Y}m^2 \right)^2 \log{\left( 1 - \frac{2m^2}{h_{\rm Y}y^2} \right)} \Biggr], 
\end{split}
\end{equation}
where $M_*$ is a cut-off scale. 
Defining a dimensionless parameter as $x \equiv y/y_c=\sqrt{h_{\rm Y}} y/\sqrt{2}m$ ($y_c\equiv \sqrt{2}Y_c)$, the total potential is rewritten as 
\begin{align}
V(\phi) &\simeq V_{\rm tree}+V_{\rm rad} \notag \\
&=m^4 \left[1+\frac{h_{\rm Y}^2 }{32 \pi^2} \left\{2 \log \left(\frac{h_{\rm Y} m^2 x^2}{M_*^2}\right)\right. \right. \notag \\
 &\quad \left. \left.+(1+x^2)^2 \log\left(1+\frac{1}{x^2}\right)+(1-x^2)^2   \log \left(1-\frac{1}{x^2}\right) \right\} \right]. 
\end{align}
The shape of $V_{\rm rad}$ is shown in Figure~\ref{fig:radpot}. 
As long as  $h_{\rm Y}^2/32 \pi^2 \ll 1$, we can approximate the potential energy with the tree-level value 
$V \simeq m^4$ and the dynamics of $y$ is governed by the 1-loop potential $V_{\rm rad}$. 
In this approximation, the slow-roll parameters are given by
\begin{equation}
\begin{split}
\epsilon &\equiv \frac{M_{\rm Pl}^2}{2} \left(\frac{V^\prime}{V}\right)^2= \frac{h_{\rm Y}^5 M_{\rm Pl}^2}{256 \pi^4 m^2}x^2\left[(x^2+1) \log \left(1+\frac{1}{x^2}\right)+(x^2-1) \log \left(1-\frac{1}{x^2}\right)\right]^2,\\
\eta &\equiv M_{\rm Pl}^2\frac{V^{\prime \prime}}{V}=\frac{h_{\rm Y}^3 M_{\rm Pl}^2}{16 \pi^2 m^2}\left[(3x^2+1)\log \left(1+\frac{1}{x^2}\right)+(3x^2-1)\log \left(1-\frac{1}{x^2}\right)\right],
\end{split}
\end{equation}
where the prime denotes the derivative with respect to $y$. 
Note that $\epsilon$ is much smaller than $|\eta|$ due to the suppression by a factor of $h_{\rm Y}^2/16\pi^2$. 
The slow-roll conditions, $\epsilon, |\eta| \ll 1$, are satisfied until $x\simeq 1 \Leftrightarrow y \simeq y_c$,  
provided that $h_{\rm Y}^{3/2}/2\sqrt{2}\pi \ll m/M_{\rm Pl}$. 
In this case, inflation ends when $y$ reaches to $y_c$ and the waterfall fields become tachyonic. 
Note that supergravity correction does not change the result significantly. 
Hereafter, we consider such a case. 

Before proceeding the discussions, we comment on the stability of this inflationary trajectory. 
In order for the above discussion to be valid, all the fields 
other than the inflaton must be heavy enough. 
Let us consider the scalar potential in supergravity, 
\begin{equation}
V_{\rm g}=e^{K/M_{\rm Pl}^2}\left[D_iW K^{i{\bar j}} D_{\bar j} W^* -\frac{3}{M_{\rm Pl}^2}|W|^2\right], 
\label{sugrap}
\end{equation}
where $K$ is the K\"ahler potential, $D_i W = \partial W/\partial \phi_i  +(\partial K/\partial \phi_i) W/M_{\rm Pl}^2$ and 
$K^{i{\bar j}}$ is the inverse matrix of $\partial^2 K/\partial \phi_i \partial \phi_{j}^*$.  
During inflation, the term in parenthesis of Eq.~\eqref{sugrap} is evaluated as $\sim 3H^2 M_{\rm Pl}^2$. 
As a result, the fields (collectively denoted $\psi$) with the canonical K\"ahler potential receive the Hubble induced masses, 
\begin{equation}
V_{\rm g} \simeq e^{|\psi|^2/M_{\rm Pl}^2} \left( 3H^2 M_{\rm Pl}^2 \right) \simeq 3 H^2 |\psi|^2 + \cdots, 
\end{equation}
and this is true for almost all the fields in the theory. 
Thus, during inflation, such fields are well stabilized at the origin of their field space.
In particular, the pseudomoduli field $\Phi$ does not lie directly on the metastable vacuum, but lies at the origin during and just after inflation.
This fact is not important for the inflationary dynamics but affects the cosmic history 
through the late time oscillation as we will see in the next section. 

We now evaluate the cosmological perturbation in this model. 
The number of $e$-folds ${\cal N}$ is calculated as 
\begin{align}
{\cal N}&= \frac{1}{M_{\rm Pl}^2}\int_{y_c}^y  \frac{V}{V^\prime}dy \simeq \frac{16\pi^2 m^2}{h_{\rm Y}^3 M_{\rm Pl}^2}\int_1^x x^\prime dx^\prime = \frac{8\pi^2 m^2}{h_{\rm Y}^3 M_{\rm Pl}^2}(x^2-1), 
\end{align}
where we have approximated with $\log[1\pm1/x^2] \simeq \pm x^{-2}$ for larger field values.\footnote{
This approximation is correct with an accuracy of ${\cal O}(1)$ in the region in which we are interested. 
}
{}From this expression, we obtain the field value evaluated ${\cal N}$ $e$-folds before the end of inflation,
\begin{equation}
y_{\cal N}= \sqrt{\frac{2}{h_{\rm Y}}}mx_{\cal N} \simeq 
 \left\{
\begin{array}{ll}
\sqrt{\dfrac{2}{h_{\rm Y}}}m & \text{for} \quad {\mathcal N} < \dfrac{8\pi^2 m^2}{h_Y^3 M_{\rm Pl}^2},\\
\dfrac{h_{\rm Y}}{2\pi} \sqrt{{\cal N} } M_{\rm Pl} &\text{for} \quad{\mathcal N} > \dfrac{8\pi^2 m^2}{h_Y^3 M_{\rm Pl}^2}.
\end{array} \right.
\end{equation}
We find that a trans-Planckian initial value can be avoided  if $\sqrt{2/h_Y}m \ll M_{\rm Pl}$ for the upper case and 
$\sqrt{{\cal N} }h_{\rm Y} < 2\pi$ for the lower case. 
We do not need to worry that the inflaton may roll off to the SUSY preserving vacuum 
when we start from such a large initial value of the inflaton field, because the pseudomoduli field 
$\Phi$ is trapped at the origin of the field space.

The primordial density perturbation generated during inflation is, then, evaluated as 
\begin{equation}
{\cal P}_{\cal R}^{1/2} \simeq \frac{1}{\sqrt{2 \epsilon}} \left( \frac{H}{2 \pi M_{\rm Pl}} \right) \simeq 
 {
 \left\{
\begin{array}{ll}
\dfrac{4 \sqrt{6 \pi}}{3}\dfrac{m^3}{h_Y^{5/2}M_{\rm Pl}^3} & \text{for} \quad {\mathcal N} < \dfrac{8\pi^2 m^2}{h_Y^3 M_{\rm Pl}^2},\\
\dfrac{2}{h_{\rm Y}}\sqrt{\dfrac{\cal N}{3}}\left(\dfrac{m}{M_{\rm Pl}}\right)^2& \text{for} \quad {\mathcal N} > \dfrac{8\pi^2 m^2}{h_Y^3 M_{\rm Pl}^2}. 
\end{array}\right.}
\end{equation}
For the COBE/WMAP normalization, ${\cal P}_{\cal R}^{1/2} \simeq  4.9\times 10^{-5}$ at 
$k=0.002 {\rm Mpc}^{-1}$ \cite{Komatsu:2010fb}, we require
\begin{equation}
\frac{m}{h_{\rm Y}^{1/2}} \simeq 5.9 \times 10^{15} {\rm GeV}  {\times \left\{
\begin{array}{cl}
\left(\dfrac{h_Y}{3 \times 10^{-3}}\right)^{1/3}&\text{for} \quad h_Y < 3\times 10^{-3},\\
\left(\dfrac{{\cal N}_{\rm COBE}}{51}\right)^{-1/4} &\text{for} \quad h_Y > 3\times 10^{-3}, 
\end{array}\right. }
\label{mscale}
\end{equation} 
where ${\cal N}_{\rm COBE}$ is the  number of $e$-folds after the COBE scale leaves the horizon. 
Note that when there is the moduli dominated era, $\mathcal{N}_{\rm COBE}$ is expressed as  
\begin{equation}
\begin{split}
\mathcal{N}_{\rm COBE} &\simeq 51 + \frac{2}{3} \log \left( \frac{V^{1/4}}{ 5.4 \times 10^{14} \, \text{GeV}} \right) + \frac{1}{3} \log \left( \frac{T_{\rm R}}{10^{10} \, \text{GeV}} \right) \\
&\qquad \qquad \qquad \qquad \qquad \qquad- \frac{1}{3} \log \left( \frac{T_{\rm dom}}{ 0.4 \, \text{GeV}} \right) + \frac{1}{3} \log \left( \frac{T_{\rm d}}{4 \, \text{MeV}} \right), 
\end{split}
\end{equation}
where $T_{\rm dom}$ and $T_{\rm d}$ {are} the cosmic temperature at the time of the moduli domination and {that at} the moduli decay, respectively.  
We will see the validity of the reference values in the next section.
The spectral tilt can be evaluated as
\begin{equation}
n_s=1-6\epsilon+2 \eta \simeq {\left\{ 
\begin{array}{ll}
1-\dfrac{h_Y^3 M_{\rm pl}^2}{2 \pi^2 m^2} \simeq 1 & \text{for} \quad  h_Y < 3\times 10^{-3},\\
1-\dfrac{1}{{\cal N}_{\rm COBE}} \simeq 0.98, & \text{for} \quad  h_Y > 3\times 10^{-3},
\end{array}\right. }
\end{equation}
and the scalar-to-tensor ratio $r$ is given by 
\begin{equation}
r =16 \epsilon \simeq { \left\{
\begin{array}{ll}
\dfrac{h_Y^{10/3}}{16 \pi^4}\left(\dfrac{h_Y^{5/6}M_{\rm pl}}{m}\right)^2 & \text{for}\quad  h_Y < 3\times 10^{-3},\\
\dfrac{h_{\rm Y}^2}{2\pi^2}\dfrac{1}{{\cal N}_{\rm COBE}}& \text{for}\quad  h_Y > 3\times 10^{-3},
\end{array} \right.}
\end{equation}
which is much smaller than $0.1$. Hereafter, we consider the case with $h_Y<3 \times 10^{-3}$ and hence 
they have not been ruled out by current observation \cite{Komatsu:2010fb}.\footnote{
The tensor perturbation is hard to be detected even by the future observations.}

\subsection{Reheating after inflation}

We here investigate the reheating temperature due to the decay of the inflaton and the waterfall fields. 
Before proceeding {with} the discussion, we express the coupling $h_{\rm Z}$ 
in terms of the observable mass parameters for later use.  
By using \eqref{mugrav}, the moduli mass \eqref{pseudomass}, the expression \eqref{h-phi} of the coupling 
$h_{\Phi}$ and the estimation \eqref{mscale}, it is given by
\begin{equation}
h_{\rm Z} \simeq 1.8 \times 10^{-3} \times \frac{1}{\sqrt{N}} \left( \frac{r_{\rm g}}{3.5} \right)^2 \left( \frac{m_{3/2}}{15 \, \text{GeV}} \right) \left( \frac{m_{\rm \Phi}}{300 \, \text{GeV}} \right) \left( \frac{m_{\tilde{g}}}{1.5 \, \text{TeV}} \right)^{-2} {\left(\frac{h_Y}{3 \times 10^{-3}}\right)^{-1/3}.}
\end{equation}
Note that the coupling $h_{\rm Z}$ has an $\mathcal{O} (10^{-3})$ value when the observable parameters are 
set for the reference values. 
Then, the messenger scale can be estimated to be $\mathcal{O} (10^{{13}}) \, \text{GeV}$. 
We can also rewrite the mass parameter $m_{\rm Z}$ in terms of the observable parameters 
by using the gaugino mass formula \eqref{soft} as
\begin{equation}
m_{\rm Z} \simeq 8.2 \times 10^{12} \, \text{GeV} \times \frac{1}{\sqrt{N}} \left( \frac{r_{\rm g}}{3.5} \right)^3 \left( \frac{m_{3/2}}{15 \, \text{GeV}} \right) \left( \frac{m_{\rm \Phi}}{300 \, \text{GeV}} \right) \left( \frac{m_{\tilde{g}}}{1.5 \, \text{TeV}} \right)^{-2},
\end{equation}
which gives the mass scale of the explicit $U(1)_R$ symmetry breaking.

\begin{figure}[tbp]
 \begin{center}
  \begin{overpic}[width=5cm,clip]{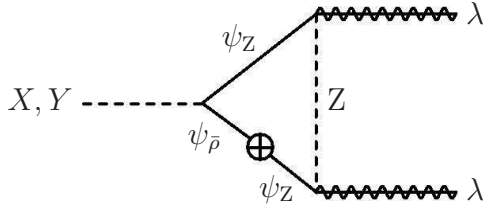}
   \put(-20,24){{$X,Y$}}
   \put(102,47){{$\lambda$}}
   \put(102,0){{$\lambda$}}
   \put(37,41){{$\psi_{\rm Z}$}}
   \put(28,15){{$\psi_{\rm \bar{\rho}}$}}
   \put(47,1){{$\psi_{\rm Z}$}}
   \put(65,24){{Z}}
  \end{overpic}
\end{center}
\caption{The decay of the waterfall field $X$ and the inflaton $Y$ into an MSSM gaugino pair. $\psi$ denotes the fermion component of the messenger multiplet.}
\label{fig:decay}
\end{figure}

After inflation, the oscillation of the waterfall fields $X \equiv \chi+\bar{\chi}$ and the inflaton dominates the universe. 
Then, the decay of these fields reheats the universe. 
These fields have $\mathcal{O}( \sqrt{h_{\rm Y}}m)$ masses as shown in Table~\ref{tab:mass} 
{and (almost) maximally mix with each other to form the mass eigenstates.}
They dominantly decay into MSSM gaugino pairs through the messenger loop if 
the decay channels to the messengers are closed, $2 h_Z > h_Y$. 
The decay diagram is given in Figure~\ref{fig:decay}.
We can estimate the decay width as
\begin{equation}
\Gamma_{\rm R} \simeq \frac{1}{2} \, \frac{N^4}{(4\pi)^5} \left( \frac{h_{\rm Y}^4 g_3^2}{h_{\rm Z}^3} \right)^2 \sqrt{h_{\rm Y}} m.
\end{equation}
{The reheating temperature is, then,}  estimated as
\begin{equation}
\begin{split}
T_{\rm R}  &\simeq \left(\frac{90}{\pi^2 g_*^{\rm R}}\right)^{1/4} \times \sqrt{ \Gamma_{\rm R} M_{\rm Pl} } \\
&\simeq {0.45} \times \frac{N^2}{(4\pi)^2} \left( \frac{\sqrt{h_{\rm Y}}}{8\pi} \right)^{1/2} \frac{h_{\rm Y}^4 g_3^2}{h_{\rm Z}^3} \, (mM_{\rm Pl})^{1/2} \\
&\simeq 5.2 \times 10^{10} \, {\rm GeV} \times N^{7/2} \left(\frac{r_g}{3.5}\right)^{-6}  \left( \frac{m_{3/2}}{15 \, \text{GeV}} \right)^{-3} \left( \frac{m_{\rm \Phi}}{300 \, \text{GeV}} \right)^{-3} \left( \frac{m_{\tilde{g}}}{1.5 \, \text{TeV}} \right)^{6} \left(\frac{h_Y}{3 \times 10^{-3}}\right)^{17/3}, 
\end{split}
\end{equation}
where $g_*^{\rm R}$ is the number of the relativistic degrees of freedom at the time of reheating and we take it as 
$g_*^{\rm R} \simeq 220$. 
The decay into the pseudomoduli field is also possible but is suppressed by the additional small Yukawa couplings. 
Since the messenger scale is estimated to be $\mathcal{O} (10^{13}) \, \text{GeV}$, the SUSY breaking sector cannot be thermalized in our scenario.

As well as the visible sector fields, gravitinos are also produced in the thermal bath at the reheating stage. 
{}From the above reheating temperature, the abundance of gravitinos is given by \cite{Kawasaki:2004yh}
\begin{equation}
\frac{\rho_{3/2}^{(\rm th)}}{s} \simeq 9.5 \times 10^{-8} \, \text{GeV} \times \left( \frac{m_{\tilde{g}}}{1.5 \, \text{TeV}} \right)^{2} \left( \frac{m_{3/2}}{15 \, \text{GeV}} \right)^{-1} \left( \frac{T_{\rm R}}{10^{10} \, \text{GeV}} \right),
\label{thab}
\end{equation}
where $s$ is the entropy density.
Note that gravitinos are also produced by other processes such as the inflaton decay, however these are sub-dominant processes if the coupling $h_{\rm Y}$ is not so small so that we can ignore these contributions to the gravitino abundance.
On the other hand, the present dark matter abundance is given by 
\cite{Komatsu:2010fb}
\begin{equation} 
\Omega_{\rm DM} h^2 \simeq 0.11, \label{dmaba}
\end{equation}
where $h\equiv H_0/(100 \ {\rm km \ sec}^{-1} {\rm Mpc}^{-1})$ and $H_0$ is the present Hubble parameter.  
Then, we have the constraint on the gravitino abundance, 
\begin{equation}
\frac{\rho_{3/2}}{s} < \frac{\rho_{\rm DM}}{s}  \simeq 4.1 \times 10^{-10}\, \text{GeV}. \label{dmabb}
\end{equation}
Since we would like to have the $\mathcal{O}(1\!-\!10) \, \text{GeV}$ gravitino mass and the $\mathcal{O}(1) \, \text{TeV}$ 
gluino mass in our scenario, we can see from Eqs.~\eqref{thab}
that gravitinos are overproduced at the time of reheating 
after inflation. Then, some late time entropy production is required for this scenario to be successful. 
Actually, this model has a possible late time entropy production mechanism, that is, the decay of the pseudomoduli field. 
As we will see, it can dilute gravitinos {sufficiently}. 
Thus, we {may} obtain the correct abundance \eqref{dmaba} of the gravitino dark matter.
We study this issue in further details in the next section.

\section{Gravitino dark matter}

In this section, we consider the cosmological evolution after {reheating and investigate the gravitino abundance, 
which is similar to the case of Ref.~\cite{Ibe:2006rc}.}
First, we discuss the oscillation of the pseudomoduli field around the SUSY breaking metastable vacuum 
and the domination in a cosmological stage. 
Next, we will investigate the decay temperature of the pseudomoduli field. 
Then, we {calculate} the gravitino abundance due to the pseudomoduli decay. 
{}From the result of the decay temperature, we also estimate the amount of the entropy production 
by the moduli {decay} and evaluate the present abundance of thermally produced gravitinos.
We find the model parameter space consistent with the constraints obtained from the analysis of these quantities.

\subsection{Moduli oscillation}

We here analyze the entropy production due to the moduli oscillation. 
As shown in the previous section, the moduli field is stabilized at the origin during inflation. 
{As the Hubble parameter decreases after inflation, the potential minimum of the pseudomoduli 
moves away from the origin. 
Then,} it starts to oscillate when the Hubble parameter becomes smaller than the moduli mass \eqref{pseudomass}. 
The temperature at that time is given by
\begin{equation}
\begin{split}
T_{\rm osc} &\simeq \left(\frac{90}{\pi^2 g_*^{\rm osc}}\right)^{1/4}  \times \sqrt{M_{\rm Pl} m_{\rm \Phi}} \\
&\simeq 1.2 \times 10^{10} \, \text{GeV} \times \left( \frac{m_{\rm \Phi}}{300 \, \text{GeV}} \right)^{1/2}.
\end{split}
\end{equation}
{Here, $g_*^{\rm osc}$ is the number of the relativistic degrees of freedom at the onset of the pseudomoduli oscillation 
and we take it as $g_*^{\rm osc}\simeq 220$.}
Since the lifetime of the moduli field is rather long as we will see, the pseudomoduli oscillation dominates the 
{energy density of the} universe. 
The temperature when the domination of the moduli oscillation occurs is given by
\begin{equation}
T_{\rm dom} \simeq 
 \left\{
\begin{array}{ll}
 \left(\dfrac{g_*^{\rm osc}}{g_*^{\rm dom}}\right)^{1/3}\left( \dfrac{|\Phi_0|}{\sqrt{3} M_{\rm Pl}} \right)^2 T_{\rm osc} & {\rm for} \quad T_{\rm R}>T_{\rm osc} , \\
  \left(\dfrac{g_*^{\rm R}}{g_*^{\rm dom}}\right)^{1/3}\left( \dfrac{|\Phi_0|}{\sqrt{3} M_{\rm Pl}} \right)^2 T_{\rm R} &  {\rm for} \quad T_{\rm R}<T_{\rm osc}, 
 \end{array}\right.
\end{equation}
{where $g_*^{\rm dom}$ is the number of the relativistic degrees of freedom at the pseudomoduli domination. 
When the pseudomoduli field decays, a large amount of entropy is produced and gravitinos are diluted.}
The dilution factor $\Delta^{-1}$ due to the pseudomoduli decay is expressed as
\begin{equation}
\Delta^{-1} \simeq \frac{T_{\rm d}}{T_{\rm dom}} \simeq 
 \left\{
\begin{array}{ll}
{\left(\dfrac{g_*^{\rm dom}}{g_*^{\rm osc}}\right)^{1/3}} \dfrac{T_{\rm d}}{T_{\rm osc}} \left( \dfrac{|\Phi_0|}{\sqrt{3} M_{\rm Pl}} \right)^{-2},&  {\rm for} \quad T_{\rm R}>T_{\rm osc},  \\
{\left(\dfrac{g_*^{\rm dom}}{g_*^{\rm R}}\right)^{1/3}} \dfrac{T_{\rm d}}{T_{\rm R}} \left( \dfrac{|\Phi_0|}{\sqrt{3} M_{\rm Pl}} \right)^{-2},
&  {\rm for} \quad T_{\rm R}<T_{\rm osc}, 
 \end{array}\right.
\end{equation}
where $T_{\rm d}$ is the decay temperature of the pseudomoduli, {which we will  estimate later.}

We here comment on the stability of the pseudomoduli oscillation. 
By using the observable mass parameters,
the expectation value of the moduli field on the metastable vacuum is rewritten as
\begin{equation}
|\Phi_0| \simeq 1.1 \times 10^{14} \, \text{GeV} \times \left( \frac{r_{\rm g}}{3.5} \right)^2 \left( \frac{m_{3/2}}{15 \, \text{GeV}} \right) \left( \frac{m_{\tilde{g}}}{1.5 \, \text{TeV}} \right)^{-1}.
\end{equation}
The messengers become tachyonic when the moduli field takes the value,
\begin{equation}
\Phi > \frac{h_{\rm Z}^2}{h_{\rm Y} h_{\rm \Phi}} \, \frac{m^2}{m_{\rm Z}} \simeq 4.0 \times 10^{14} \, {\rm GeV} \times \left(\frac{m_{3/2} }{15 \, {\rm GeV}}\right) \left(\frac{m_{\tilde g}}{1.5 \, {\rm TeV}}\right)^{-1}. 
\end{equation}
In order for the field configuration to settle down finally to the metastable vacuum, 
we require the following condition on the gaugino-to-scalar mass ratio:
\begin{equation}
r_{\rm g} \lesssim 4.5.
\end{equation} 
Compared to the case in Ref.~\cite{Ibe:2006rc}, the amplitude of the oscillation is rather small. 
Then, the abundance of {gravitinos produced at the time of reheating}
is comparable to that of gravitinos due to the moduli decay.

\subsection{Moduli decay and DM abundance}

Next, we analyze the decay temperature of the pseudomoduli field and the abundance of the gravitino dark matter. 
The pseudomoduli can decay into both the visible-sector particles and gravitinos through the messenger loops 
which are encoded in the effective interactions of the pseudomoduli field with the visible-sector fields. 
We consider the case {that} the pseudomoduli field can decay into two Higgs bosons, $i.e.$, 
$m_{\rm \Phi} > 2m_h$. 
We will see the case {that} the moduli mass is small and the field cannot decay into two Higgs bosons in Appendix.

The interaction Lagrangian of the pseudomoduli field with the scalar components of the visible-sector superfields 
can be extracted from the moduli dependence of the soft scalar mass squared, $m_{\tilde{f}}^2(\Phi)$, which is given by
\begin{equation}
m_{\tilde{f}}^2(\Phi) = {\sum_i} C_2^i \left( \frac{g_i^2}{16\pi^2} \right)^2 \left[ \frac{h_{\rm Y}h_{\rm \Phi}^2}{h_{\rm Z}^2} \, \frac{\mu^4}{m^2} + \frac{3}{4} \, \frac{h_{\rm Y}^2 h_{\rm \Phi}^3}{h_{\rm Z}^4} \, \frac{\mu^4 m_{\rm Z}}{m^4} \left(\Phi + \Phi^\dagger \right) + \cdots \right],
\end{equation}
where the first term is the leading order of the scale $m_Z$ and leads to the soft masses shown in section 2. 
The second term is the next-to-leading order one. 
There exist higher order terms {as well}, but we do not write them explicitly. 
The effective interactions of the moduli field can be extracted from this expression  as
\begin{equation}
\begin{split}
\mathcal{L}_{\tilde{f}} &=  \frac{\partial m_{\tilde{f}}^2 (\Phi)}{\partial \Phi} \, \Phi \tilde{f} \tilde{f}^\dagger + {\rm h.c.} \\
&\simeq \frac{3}{4} \, {\sum_i}C_2^i \left( \frac{g_i^2}{16\pi^2} \right)^2 \frac{h_{\rm Y}^2h_{\rm \Phi}^3}{h_{\rm Z}^4} \, \frac{\mu^4 m_{\rm Z}}{m^4} \, \Phi \tilde{f} \tilde{f}^\dagger + {\rm h.c.}
\label{modint}
\end{split}
\end{equation}
The pseudomoduli field dominantly decays into two Higgs bosons through the messenger loop. 
{Other scalar fields are too heavy for the pseudomoduli field to decay into them.}
The decays into the standard model gauge bosons are sub-dominant because the effective interactions 
of the pseudomoduli field with the gauge bosons (shown in Appendix) are suppressed by a 1-loop factor more than 
the interaction given above. 
The decay width is estimated from this effective interaction \eqref{modint}  as
\begin{equation}
\Gamma_{\rm H} \simeq \frac{24 \pi^3}{N^2} \, \frac{x_{\rm H}^2}{x_{\rm g}^2} \, \frac{1}{h_{\rm \Phi}^4} \left( \frac{m_{\rm \Phi} m_{\tilde{g}}}{M_{\rm Pl}m_{3/2}} \right)^2 m_{\rm \Phi},
\label{modde}
\end{equation}
where $x_{\rm H}$ and  $x_{\rm g}$  are {defined as} 
\begin{align}
x_{\rm H} &\equiv \frac{g_2^4}{(4\pi)^4} \cdot \frac{3}{4} + \frac{g_{\rm Y}^4}{(4\pi)^4} \cdot \frac{5}{3} \cdot \frac{1}{4} \simeq 6 \times 10^{-6}, \\
x_{\rm g} & \equiv \frac{g_3^2}{(4\pi)^2} \simeq 9.4 \times 10^{-3}.
\end{align}
We have used the expressions of the observable mass parameters such as the gluino mass $m_{\tilde{g}}$. 
The moduli decay {produces large entropy in the universe}
and the temperature after the decay can be estimated from the decay width \eqref{modde} as
\begin{equation}
\begin{split}
T_{\rm d} &\simeq \sqrt{\Gamma_{\rm H} M_{\rm Pl}} \\
&\simeq {4.4} \, \text{MeV} \times \left( \frac{r_{\rm g}}{3.5} \right)^{-2} \left( \frac{m_{\tilde{g}}}{1.5 \, \text{TeV}} \right)^3 \left( \frac{m_{3/2}}{15 \, \text{GeV}} \right)^{-1} \left( \frac{m_{\rm \Phi}}{300 \, \text{GeV}} \right)^{-1/2}.
\label{dt}
\end{split}
\end{equation}
The temperature is required to be above $2 \, \text{MeV}$ so that the standard BBN properly occurs.\footnote{{
Baryogenesis must take place before the BBN. 
The required baryon-to-photon ratio is $n_B/n_\gamma \sim 6 \times 10^{-10}$ \cite{Komatsu:2010fb}.  
In the case that the moduli oscillation dominates the energy density of the universe, larger baryon asymmetry must be generated 
in order to take into account the entropy production from the moduli decay.}}
This requirement constrains on the parameters in our model.

We next calculate the number density of gravitinos from the pseudomoduli decay. 
The decay into the longitudinal mode of gravitino (would-be goldstino) is the dominant process. 
The interaction Lagrangian with the longitudinal mode can be read off from the effective K\"ahler potential \eqref{keffmod},
\begin{equation}
\mathcal{L}_{3/2} \simeq - \frac{N}{(16\pi)^2 } \, \frac{h_{\rm Y}h_{\rm \Phi}^4}{h_{\rm Z}^2} \, \left( \frac{\mu}{m} \right)^2  \Phi^\dagger  \bar{\psi}_{3/2} \psi_{3/2} + \rm h.c.,
\end{equation}
where $\psi_{3/2}$ denotes the gravitino. 
Then, the partial decay width of the pseudomoduli field can be calculated as
\begin{equation}
\Gamma_{3/2} \simeq \frac{1}{192\pi} \left(\frac{m_{\rm \Phi}^2}{M_{\rm Pl}m_{3/2}} \right)^2 m_{\rm \Phi},
\label{pdw}
\end{equation}
where we have used the expressions of the observable mass parameters. 
Compared to the {total} decay width \eqref{modde}, the {decay} width {into gravitinos} \eqref{pdw} is small 
by the quartic dependence of the coupling $h_{\rm \Phi}$. 
The number density of gravitinos is expressed as
\begin{equation}
\frac{n_{3/2}}{s} = \frac{3}{4} \, \frac{T_{\rm d}}{m_{\rm \Phi}} \, B_{3/2} \times 2,
\end{equation}
where $B_{3/2}(= \Gamma_{3/2}/\Gamma_H)$ is the branching fraction into two gravitinos and $s$ is the entropy density of the universe. The branching fraction is given by
\begin{equation}
B_{3/2} \simeq \frac{1}{18} \, \frac{N^2}{(4\pi)^4} \, \frac{x_{\rm g}^2}{x_{\rm H}^2} \, h_{\rm \Phi}^4  \left( \frac{m_{\rm \Phi}}{m_{\tilde{g}}} \right)^2.
\end{equation}
Then, the density parameter of non-thermally produced gravitinos is estimated as
\begin{equation}
\begin{split}
\Omega_{3/2}^{(\rm d)} \, h^2 &\simeq {0.033} \times \left( \frac{r_{\rm g}}{3.5} \right)^2 \left( \frac{m_{\rm \Phi}}{300 \, \text{GeV}} \right)^{9/2} \left( \frac{m_{\tilde{g}}}{1.5 \, \text{TeV}} \right)^{-3}. 
\label{nonthgrav}
\end{split}
\end{equation}
This quantity is required to be below the present dark matter abundance, $\Omega_{\rm DM} \, h^2 \simeq 0.11$. This gives another constraint on the parameters in our model.

Since we have calculated the decay temperature of the moduli field, 
we can estimate the dilution factor $\Delta^{-1}$ due to the moduli domination before the decay. 
We here require that the reheating temperature is smaller than the temperature at the onset of the 
pseudomoduli oscillation, $T_{\rm R} \lesssim T_{\rm osc}$.
The reason is as follows. 
If the reheating temperature after inflation is too high, the abundance of thermally produced gravitinos is too large 
to explain the present dark matter abundance even if we include the effect of the dilution {by the pseudomoduli decay.\footnote{If the reheating temperature is too low, we cannot neglect the direct gravitino production \cite{Endo:2007sz}. We do not consider such a case.}
Then, from the condition $T_{\rm R} \lesssim T_{\rm osc}$, the constraint on the coupling 
$h_{\rm Y}$ can be expressed in terms of the observable parameters as
\begin{equation}
h_{\rm Y} \lesssim 2.2 \times 10^{-3} \times \frac{1}{N^{21/34}} \left( \frac{r_{\rm g}}{3.5} \right)^{18/17} \left( \frac{m_{3/2}}{15 \, \text{GeV}} \right)^{9/17} \left( \frac{m_{\rm \Phi}}{300 \, \text{GeV}} \right)^{21/34} \left( \frac{m_{\tilde{g}}}{1.5 \, \text{TeV}} \right)^{-18/17}.
\label{hphi}
\end{equation}
Note that $h_{\rm Y}$ is an $\mathcal{O} (10^{-3})$ quantity as assumed above. 
The resulting abundance of gravitinos {produced at the time of reheating} is given by
\begin{equation}
\Omega_{3/2}^{(\rm th)} \, h^2 \simeq 0.016 \times {\left(\frac{g_*^{\rm dom}}{g_*^{\rm osc}}\right)^{1/3}} \left( \frac{r_{\rm g}}{3.5} \right)^{-6} \left( \frac{m_{\rm \Phi}}{300 \, \text{GeV}} \right)^{-1/2} \left( \frac{m_{\tilde{g}}}{1.5 \, \text{TeV}} \right)^{7} \left( \frac{m_{3/2}}{15 \, \text{GeV}} \right)^{-4},
\label{thgrav}
\end{equation}
which is required to be below the present dark matter abundance. This gives further constraint on the parameters in our model.

\begin{figure}[t]
\begin{tabular}{cc}
\includegraphics[width=80mm]{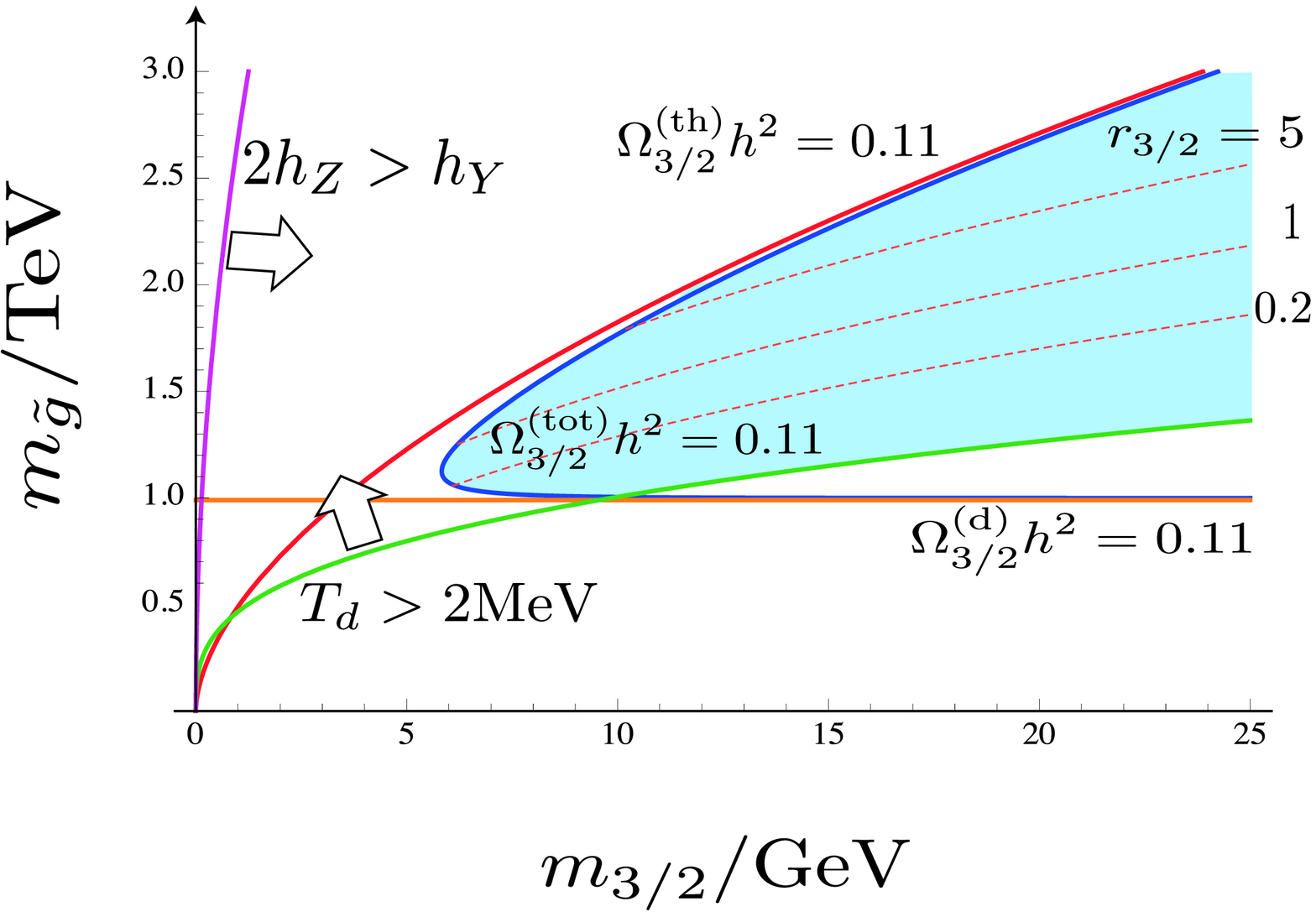}
 &
 \includegraphics[width=80mm]{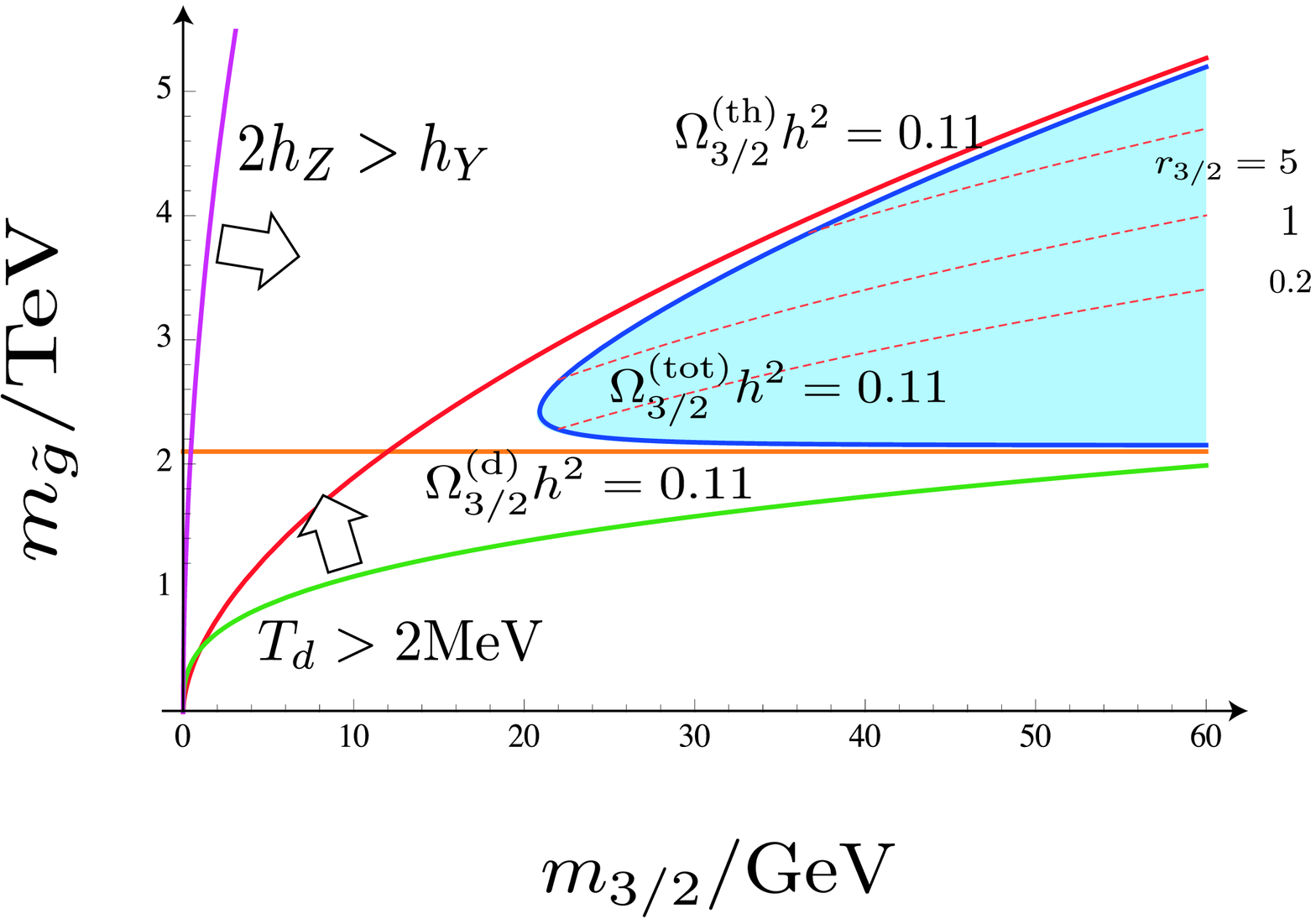}
\end{tabular}
\vspace{-0.3cm}
\caption{The allowed region (shaded region) of the gravitino mass (horizontal axis) and the gluino mass (longitudinal axis). In the left and right figures, the moduli mass is taken to be $300 \, \text{GeV}$ and $500 \, \text{GeV}$, respectively. The gaugino-to-scalar mass ratio is $r_{\rm g} = 3.5$.}
\label{fig:gravconst}
\end{figure}

\begin{figure}[t]
\begin{tabular}{cc}
\includegraphics[width=80mm]{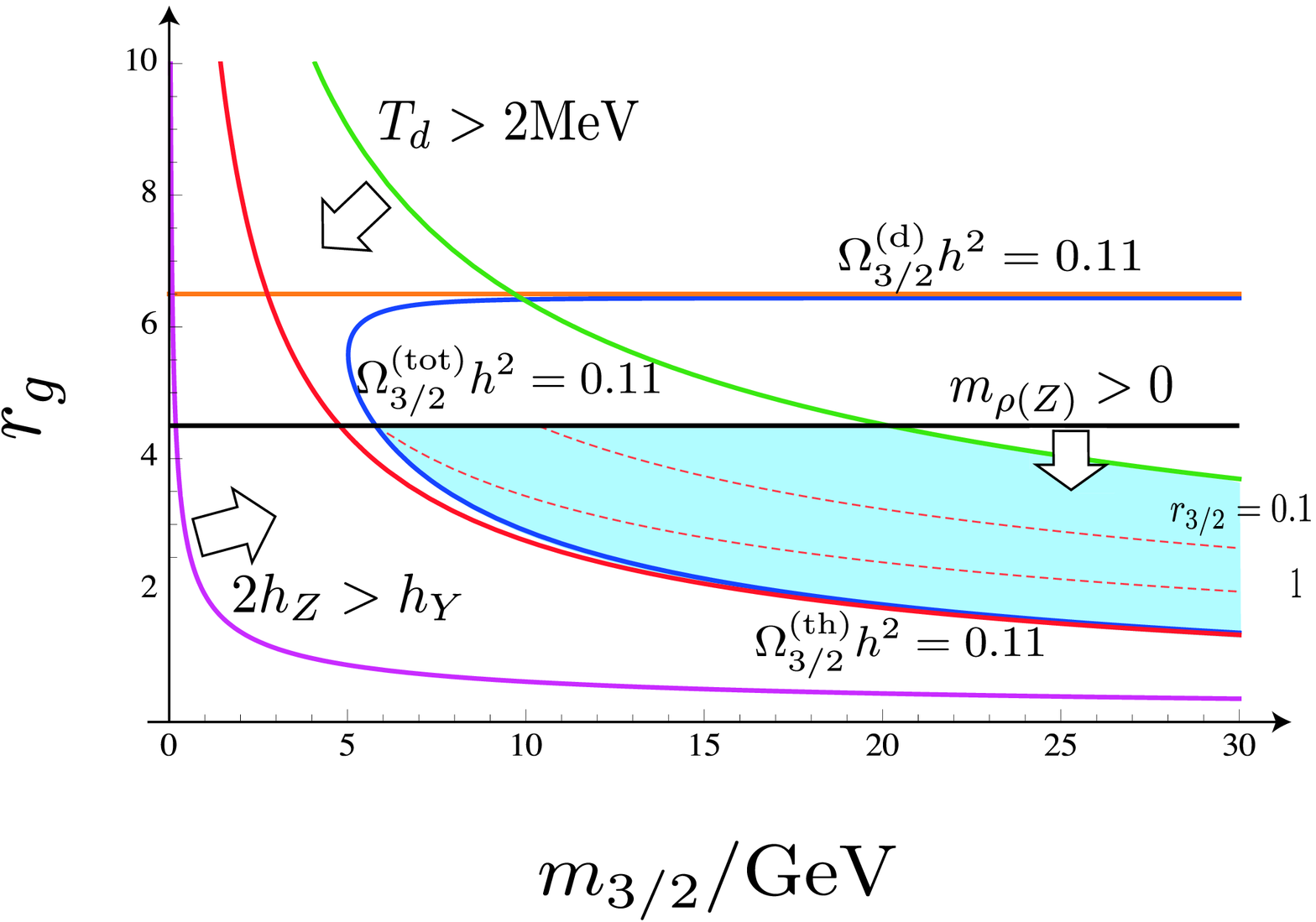}
 &
 \includegraphics[width=80mm]{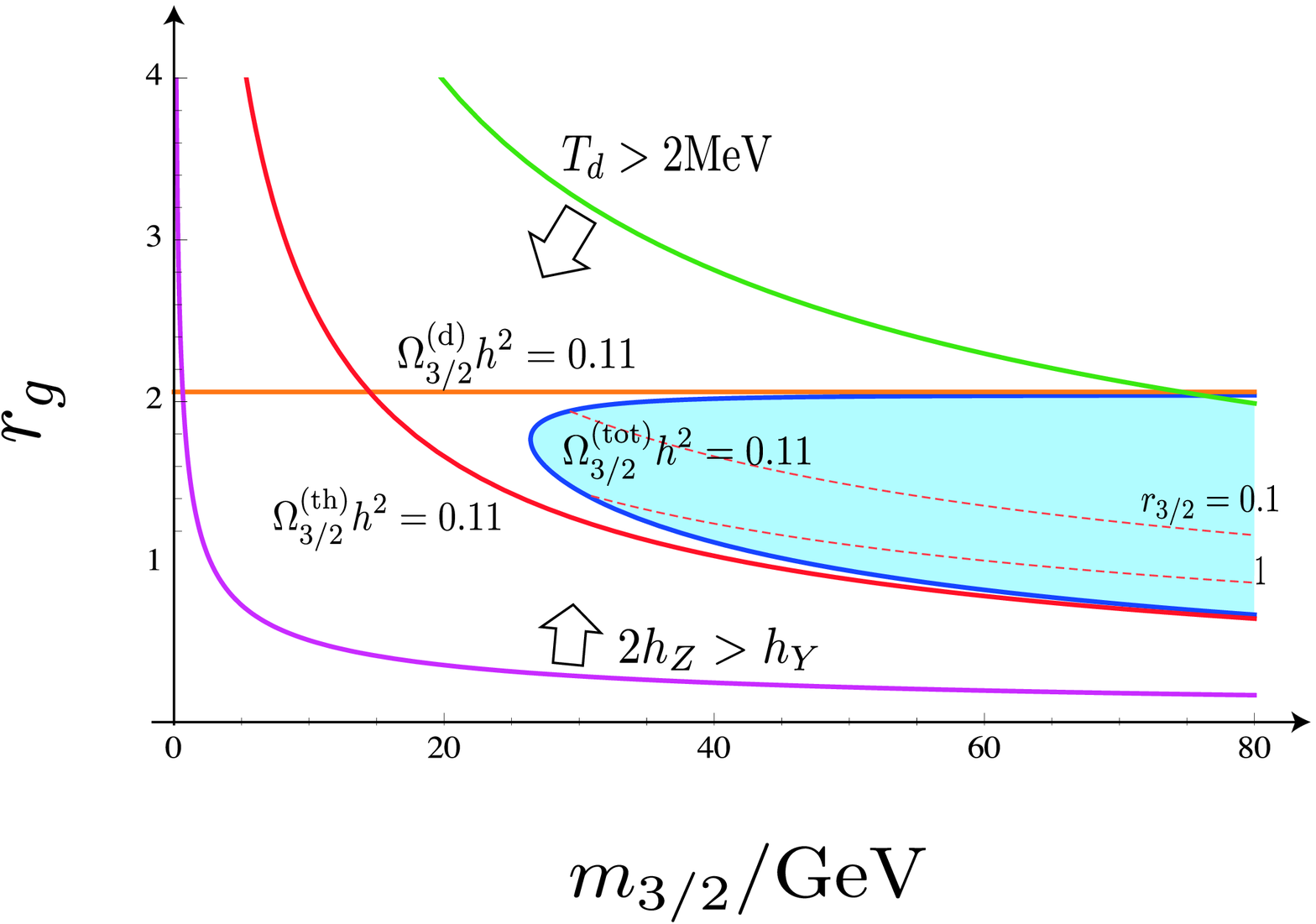}
\end{tabular}
\vspace{-0.3cm}
\caption{The allowed region (shaded region) of the gravitino mass (horizontal axis) and the gaugino-to-scalar mass ratio (longitudinal axis). In the left and right figures, the moduli mass is taken to be $300 \, \text{GeV}$ and $500 \, \text{GeV}$ respectively. The gluino mass is $1.5 \, \text{TeV}$.}
\label{fig:gravconst3}
\end{figure}

Now, we see the existence of the parameter space where all the constraints are satisfied. 
Figure~\ref{fig:gravconst} shows the allowed region (shaded region) of the gravitino mass (horizontal axis) 
and the gluino mass (longitudinal axis) which satisfies the constraints from the decay temperature \eqref{dt}
and the gravitino abundance \eqref{nonthgrav}, \eqref{thgrav}. 
We also show the parameter region where the total gravitino abundance corresponds to the present dark matter abundance.
Here, the moduli mass is taken to be $300 \, \text{GeV}$ (left figure) or $500 \, \text{GeV}$ (right figure). 
The gaugino-to-scalar mass ratio is set to be $r_{\rm g} = 3.5$ in both cases. 
The constraint $2 h_{\rm Z} > h_{\rm Y}$ is presented
so that the inflaton and the waterfall fields {would} not decay into the messengers, which {would change} our thermal history. 
We can see from the left figure that the thermally produced gravitino abundance and the decay temperature put strong 
constraints on the gluino mass. 
For the region of the small gravitino mass, the abundance of non-thermally produced gravitinos has an important effect on 
the allowed gluino mass region. On the other hand, in the right figure, the abundance of  gravitinos 
from the pseudomoduli decay puts a strong constraint on the allowed region of the gluino mass 
and the constraint from the decay temperature becomes weak. 
The allowed gravitino mass is always larger than $\mathcal{O} (10) \, \text{GeV}$. {It may be possible for the gravitino mass to be much larger than  $\mathcal{O} (10) \, \text{GeV}$, but it 
is not favored because the flavor-changing neutral current would not be suppressed in such a case. 
Then, we conclude that the gravitino mass of the order of $10 \, \text{GeV}$ is favored for our model.}
We define the ratio of the gravitino abundance produced {at the time of reheating and the pseudomoduli decay,}
$r_{3/2}\equiv \Omega_{3/2}^{(\rm th)}/\Omega_{3/2}^{(\rm d)}$. 
As the gluino mass is decreasing, the ratio also decreases and the abundance of  gravitinos from the pseudomoduli decay 
becomes dominant.

Figure~\ref{fig:gravconst3} shows the allowed region (shaded region) of the gravitino mass (horizontal axis) 
and the gaugino-to-scalar mass ratio (longitudinal axis). 
In the left {(right)} figure, the moduli mass is taken to be $300 \, \text{GeV}$ {($500 \, \text{GeV}$)}. 
{Here we take the gluino mass as $1.5 \, \text{TeV}$. }
{In the left figure, the upper bound for the mass ratio comes from the constraint that the moduli oscillation is stable, 
in other words,  the messenger fields do not become tachyonic during the moduli oscillation. 
The constraint from the gravitino abundance from the pseudomoduli decay gives a weaker bound.}
On the other hand, in the right figure, the gravitino abundance {from the pseudomoduli decay} 
gives the stronger constraint on the mass ratio. 
We can also see that the abundance of gravitinos {produced at the time of reheating} 
gives a lower bound for the mass ratio in both the left and right figures. 
As the mass ratio is decreasing, the ratio $r_{3/2}$ increases and the abundance of thermally produced gravitinos 
becomes dominant.

As discussed in Ref.~\cite{Ibe:2006rc}, if the Bino is the next to lightest supersymmetric particle (NLSP), 
the moduli field can decay into two Binos. In this case, 
these Binos decay into gravitinos later, which breaks BBN. 
However, in the case that the moduli, gluino and gravitino masses take the reference values 
and the gaugino-to-scalar mass ratio is $r_{\rm g} = 3.5$, the Bino mass is around $230 \, \text{GeV}$. 
Then, the moduli field cannot decay into two Binos and hence this problem does not occur.
{In conclusion, there is a model parameter space where this scenario can be successful. }

\section{Conclusion}

In this paper, we have proposed an inflationary {universe} scenario 
with gauge-mediated SUSY breaking. 
The higher mass scale in the model is set for the inflationary scale, 
and the lower mass scale corresponds to the SUSY breaking scale to give the correct MSSM soft masses 
by direct gauge mediation. 
After inflation, the metastable SUSY breaking vacuum is chosen naturally. 
We have analyzed the reheating stage of the model. 
We have studied the pseudomoduli oscillation and its decay as was studied in Ref.~\cite{Ibe:2006rc}. 
We find that the pseudomoduli oscillation dominates the universe after reheating 
and dilute thermally produced gravitinos. 
Gravitinos are also produced by the pseudomoduli decay. 
We find a model parameter space consistent with the gravitino dark matter. 

We would like to comment on the two issues remained in our model. 
One is the cosmic string problem. 
We here take a hybrid inflation model as a concrete realization of our cosmological scenario, 
which suffers from the cosmic string problem. 
We have set the coupling constant $h_Y$ as ${\cal O}(10^{-3})$ in order to suppress the gravitino abundance 
produced directly from the inflaton decay. 
Then, the cosmic strings may be produced at the end of inflation because the waterfall fields 
break the $U(1)_1$ symmetry 
and their tension $G\mu$ is calculated as $G\mu \sim 2 \pi m^2/h_Y$. 
Recent observation of the CMB constrains the value of the cosmic string tension, $G\mu<(2-7)\times 10^{-7}$ 
\cite{Bevis:2007gh}.  
In the case of the standard hybrid inflation with the logarithmic potential, the constraint on the coupling constant is 
expressed as $h_Y \ll 10^{-5}$,  
which is conflicted with the typical value we have applied.\footnote{
Another constraints on the  cosmic string tension from the gravitational waves (GWs) has been reported 
recently \cite{vanHaasteren:2011ni}.} 
Therefore, the cosmic  string problem is severe in our present model. 
However, we can modify our model to avoid this problem \cite{Jeannerot:2006jj} by 
changing the vacuum structure \cite{Preskill:1992bf} or replace the model from the standard hybrid inflation model
to the shifted or smooth hybrid inflation model \cite{Jeannerot:2000sv}. 
By applying them, the cosmic string would become unstable or be diluted enough not to affect the CMB or GW observation. 
We leave the categorization of these modifications for a future work.

{Another issue is baryon asymmetry in the present universe in our scenario.}
Since there is a large dilution factor $\Delta^{-1} \simeq 10^{-3}$, 
we have to produce a large amount of baryon asymmetry before the pseudomoduli domination. 
One candidate is the Affleck-Dine mechanism \cite{Affleck:1984fy}.\footnote{{
The suppression  of the net baryon asymmetry due to the nonzero temperature effect \cite{Fujii:2001zr}
can be avoided if we consider the multiple field case \cite{Kamada:2008sv}.}}
It may be interesting to find a way to generate baryon asymmetry in the SUSY breaking sector, 
which is also left for the future study.}


\section*{Acknowledgments}

We would like to thank M. Endo, M. Ibe, K.-I. Izawa, Y. Ookouchi, O. Seto, T. Suyama, F. Takahashi 
and J. Yokoyama for discussions.
This work was partially supported by JSPS through research fellowships
(K.K.) and the Grant-in-Aid for the Global COE Program ``Global Center
of Excellence for Physical Sciences Frontier'' from the Ministry of Education, Culture, Sports, Science and 
Technology (MEXT) of Japan.   
This work was also supported by the Grant-in-Aid for the Global COE Program ``The Next Generation of Physics, Spun from Universality and Emergence'' from the MEXT of Japan.
Y.N. is grateful to Institute for the Physics and Mathematics of the Universe for their hospitality during the course of this work.

\setcounter{equation}{0}
\renewcommand{\theequation}{A.\arabic{equation}}

\section*{Appendix}

We can also consider the case where the moduli mass is smaller than the twice the Higgs mass, $m_{\rm \Phi} < 2m_h$, and the moduli field decays dominantly into a gluon pair. The interaction with the standard model gauge boson can be extracted from the moduli dependence of the running gauge coupling $g_3(\Phi)$. The result is given by
\begin{equation}
\mathcal{L}_{\rm F} = -\frac{1}{4} \, \frac{g_3^2}{(4\pi)^2} \, \frac{h_{\rm Y} h_{\rm \Phi}}{h_{\rm Z}^2} \, \frac{m_{\rm Z}}{m^2} \, \Phi F_{\mu \nu}F^{\mu \nu} + {\rm h.c.}
\end{equation}
Then, the decay rate of this process is calculated to be
\begin{equation}
\Gamma_{\rm g} \simeq \frac{x_{\rm g}^2}{{8}\pi} \left(\frac{h_{\rm Y} h_{\rm \Phi}}{h_{\rm Z}^2} \right)^2 \left(\frac{m_{\rm Z} m_{\rm \Phi}}{m^2} \right)^2 m_{\rm \Phi}.
\end{equation}
As in the case discussed in the main text, the temperature after the moduli decay can be calculated from the above decay rate as
\begin{equation}
T_{\rm d} \simeq \frac{1}{\sqrt{{24}\pi}} \left( \frac{m_{\rm \Phi}}{M_{\rm Pl}} \right)^{1/2} \frac{m_{\rm \Phi} m_{\tilde{g}}}{m_{3/2}} {\simeq 2.3 \, {\rm MeV} \times \left(\frac{m_\Phi}{10 \,{\rm GeV}}\right)^{3/2} \left(\frac{m_{\tilde g}}{1 \,{\rm TeV}}\right) \left(\frac{m_{3/2}}{1 \,{\rm MeV}}\right)^{-1}}.
\end{equation}
The temperature is required to be above $2 \, \text{MeV}$ so that the standard BBN properly occurs. This requirement constrains on the parameters in our model just as in the case in the main text. By the way, using the partial decay width \eqref{pdw}, the branching fraction of the decay into gravitinos is given by
\begin{equation}
B_{3/2} \simeq \frac{1}{{8}} \left(\frac{m_{\rm \Phi}}{m_{\tilde{g}}} \right)^2.
\end{equation}
Then, the present energy density-to-entropy ratio of gravitinos is estimated as
\begin{equation}
\frac{\rho^{(\rm d)}_{3/2}}{s} \simeq \frac{1}{{16}} \sqrt{\frac{3}{{8}\pi}} \left( \frac{m_{\rm \Phi}}{M_{\rm Pl}} \right)^{1/2} \frac{m_{\rm \Phi}^2}{m_{\tilde{g}}} {\simeq 4.4 \times 10^{-12} \, {\rm GeV}  \times \left(\frac{m_\Phi}{10 \,{\rm GeV}}\right)^{5/2} \left(\frac{m_{\tilde g}}{1 \,{\rm TeV}}\right)^{-1}}.
\end{equation}
{On the other hand, although gravitinos produced at the time of reheating is diluted by the pseudomoduli decay, 
their present energy density-to-entropy ratio is still large, 
\begin{equation}
\frac{\rho^{(\rm th)}_{3/2}}{s} \simeq 22 \, {\rm GeV} \times \left(\frac{g_*^{\rm dom}}{g_*^{\rm osc}}\right)^{1/3} \left(\frac{r_g}{3.5}\right)^{-4}  \left(\frac{m_\Phi}{10 \,{\rm GeV}}\right)^{3/2} \left(\frac{m_{\tilde g}}{1 \,{\rm TeV}}\right)^5 \left(\frac{m_{3/2}}{1 \,{\rm MeV}}\right)^{-4}. 
\end{equation}
Thus, from \eqref{dmabb}, we can see that the total present energy density-to-entropy ratio of gravitinos, 
$(\rho_{3/2}^{\rm (d)}+\rho_{3/2}^{\rm (th)})/s$, is so large that they overclose the universe. 
As a result, this scenario cannot be compatible with the present universe.}

%
%

\end{document}